\documentclass[aps,prl,reprint,superscriptaddress,showpacs,amsmath,amssymb,longbibliography,lengthcheck]{revtex4-1}

\usepackage[dvipdfmx]{graphicx}
\usepackage{color}
\usepackage{txfonts}

\begin{document}

\title{Novel approach to excitation spectrum from correlated ground state}

\author{Takaharu Otsuka}
\affiliation{Department of Physics, University of Tokyo, Hongo, Bunkyo-ku, Tokyo 113-0033, Japan}
\affiliation{Center for Nuclear Study, University of Tokyo, Hongo, Bunkyo-ku, Tokyo 113-0033, Japan}
\affiliation{RIKEN Nishina Center, 2-1 Hirosawa, Wako, Saitama 351-0198, Japan}  
\affiliation{Instituut voor Kern- en Stralingsfysica, KU Leuven, B-3001 Leuven, Belgium}
\affiliation{National Superconducting Cyclotron Laboratory,Michigan State University, East Lansing, Michigan 48824, USA}
\author{Tomoaki Togashi}
\affiliation{Center for Nuclear Study, University of Tokyo, Hongo, Bunkyo-ku, Tokyo 113-0033, Japan}
\author{Noritaka Shimizu}
\affiliation{Center for Nuclear Study, University of Tokyo, Hongo, Bunkyo-ku, Tokyo 113-0033, Japan}
\author{Yutaka Utsuno}
\affiliation{Advanced Science Research Center, Japan Atomic Energy Agency, Tokai, Ibaraki 319-1195, Japan}
\author{Toshio Suzuki}
\affiliation{Department of Physics and Graduate School of Integrated Basic Sciences, 
College of Humanities and Sciences, Nihon University, Sakurajosui, Setagaya-ku, Tokyo 156-8550, Japan}
\affiliation{National Astronomical Observatory of Japan, Mitaka, Tokyo 181-8588, Japan}

\begin{abstract}
\noindent
A novel approach to obtain the excitation spectrum of nuclei is presented as well as its proof-of-principle.  
The Monte Carlo Shell Model is extended so that the excitation spectrum can be calculated from its ground state with full of correlations.  This new methodology is sketched with the example of E1 excitations from the nucleus  $^{88}$Sr in comparison to experiment.  From the $B(E 1; 0^+_1\rightarrow1^-_1)$ value, the photoabsorption cross section is calculated, with the Giant Dipole and Pygmy Dipole Resonances in agreement with experiment.  Applications to $^{90}$Sr and $^{90,93}$Zr are shown with similar characteristics.  The possible relevance to the transmutation of long-lived fission products is discussed. 
\end{abstract}
% In this example, twelve single-particle orbits including more than two Harmonic Oscillater shells are taken.    


\maketitle

The response of atomic nuclei to various excitation modes such as electric dipole (E1) excitation \cite{ring_schuck,harakeh_book,bortignon_book}, is the subject of much interest in many-body physics but is also of relevance to many fields of science.  
It involves structure aspects coming from the initial state and excitation aspects related to final states.  
Thus, this subject is more complex than the usual nuclear-structure problems which are already quite complicated.
We challenge this subject by extending the Monte Carlo Shell Model (MCSM) so that various many-body correlations can be incorporated into the initial state \cite{mcsm_review1,mcsm_ptep}, and the excitation can be described by developing a novel method which makes use of such a fully correlated initial state.  
This Letter presents the formulation of this method as well as its proof-of-principle in terms of E1 excitations of Sr and Zr isotopes, some of which are crucial long-lived fission products in the transmutation of radioactive waste.

The E1 excitations exhibit a variety of interesting physics manifestations, including the Giant Dipole Resonance (GDR) and 
Pygmy Dipole Resonance (PDR).  The GDR is a typical collective mode and understood as a motion of protons as a whole against neutrons as a whole, with a high excitation energy due to the resistance from the strong proton-neutron attraction  \cite{harakeh_book,bortignon_book}.  The PDR has been studied more recently as low-lying E1 excitations with certain strengths though much weaker than GDRs \cite{aumann2013,savran2013}.   In this Letter, we shall discuss how the GDR and PDR appear in the general framework to be presented.
 
%%%  Figure 1   
\begin{figure}[tb]
  \includegraphics[width=8.0cm]{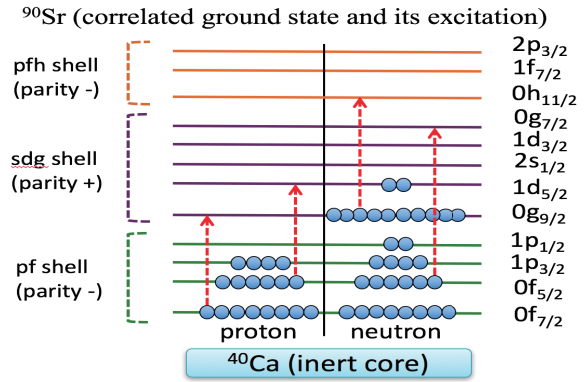}
  \caption{(Color online)  Schematic picture of the present approach.  A wide model space is taken, the correlated ground state is obtained by the MCSM, and E1 excitations (red dashed vertical lines) are calculated.  The occupation pattern shown is an example. }
  \label{fig:Sr_space}
\end{figure}

The E1 excitation has been studied in terms of various types of RPA calculations \cite{ring_schuck}.  In the present work, we develop a completely different approach by extending shell-model (SM; also called Configuration Interaction) calculations from the structure studies to the excitation spectra.  The initial state of the excitation is usually the ground state, which is assumed also in this work.  The present SM approach uses single-particle orbits of the  Harmonic Oscillator (HO) potential.  The E1 transition involves two orbits of opposite parity, changing HO quanta by one, {\it i.e.}, connecting to neighboring HO shells.  The E1 excitations from lowest single-particle orbits are blocked because final single-particle states are also occupied.  Thus, E1 excitations, including GDR, can be described well by considering the highest part of the inert core, valence orbits, and their neighboring higher orbits.
In the SM calculation, active single-particle orbits make up the model space.  
For Sr and Zr isotopes to be studied, the appropriate and still tractable model space can be the one shown in Fig.~\ref{fig:Sr_space}: the full $pf$-shell, the full $sdg$-shell, and the lower part of the $pfh$-shell.  
The nucleons are assumed to fill all orbits of the $^{40}$Ca core, forming the small inert core.  The remaining nucleons occupy mainly lower orbits of this model space, contributing to E1 excitations.  This model space is much wider than the one taken in normal SM calculations.

A superposition of various configurations produces correlation energies, giving rise to the ground state.  This is nothing but the eigensolution of the many-body Schr\"odinger equation.  This eigensolution is obtained by diagonalizing the Hamiltonian matrix, which is not feasible for the present examples with traditional shell-model methods because of the dimension of the matrix ($\sim10^{37}$).  On the other hand, such eigensolutions can still be obtained by the MCSM, with the eigenstate expressed as
%%%%%%%  eq.(1)
\begin{equation}
| \, \Psi ({\bf D}) \rangle = \sum _{n=1}^{N_B}  \, c_n \, P^{J^{\pi}} \,|\, \phi^{(n)}  \rangle,
\label{eq:eigenstate}
\end{equation}
where $P^{J^{\pi}}$ is the projection operator onto the spin-parity $J^{\pi}$,  $|\, \phi^{(n)}  \rangle$ denotes the n-th MCSM basis vector (a deformed Slater determinant as described below) with $N_B$ and $c_n$ implying, respectively, the number of such basis vectors and the amplitude.  
Here, ${\bf D}$ stands for a set of matrices $D^{(n)} \, (n=1, 2, ...)$, the matrix elements of which appear in the Slater determinant expressed as the direct product, 
%%%%%%%  eq.(2)
\begin{equation}
|\, \phi^{(n)}  \rangle 
= \prod _{\alpha=1}^{N_p}  \, \left( \,  \sum _{i=1}^{N_s}  \, a^{\dagger}_{i} \, D^{(n)}_{i \alpha} \right) |0 \rangle,
\label{eq:basisvector}
\end{equation}
where $N_p$ ($N_s$)  is the number of valence particles (the number of single-particle states), $a^{\dagger}_{i}$ means the creation operator of the i-th original single-particle state, and $|0 \rangle$ denotes the inert core.  
Here, $D^{(n)}_{i \alpha}$ are the amplitudes to expand the $\alpha$-th deformed single-particle state 
(forming the Slater determinant) by the original single-particle states with index i,  
with respect to the n-th MCSM basis vector.  
The matrices  ${\bf D}$, through which the eigenwavefunction is specified, are determined by combining stochastic, 
variational and diagonalization procedures \cite{mcsm_review1,mcsm_ptep}.  

Once the ground state is fixed, we move on to the excitation from this 
ground state with many correlations.  
In the case of a nucleus with an even number of protons ($Z$) and an even number of neutrons ($N$), the ground state is a 0$^+$ state, while one has to consider many excited 1$^{-}$ states.  
It is not a good idea to prepare all 1$^{-}$ eigenstates and calculate E1 transitions to them, because the number of such 1$^{-}$ states can be prohibitively large and most of them are not or only weakly linked to the ground state through E1 transitions.  We, however, do not need them all.  What is needed is the E1 strength distribution.
This viewpoint leads us to a novel idea that we can let the E1 transition specify a Hilbert subspace out of the whole 1$^{-}$ space, and we can investigate E1 excitations from a specific state within this subspace.   We now sketch this idea.      

We consider the action of the E1 operator
%%%%%%  eq.(3)
\begin{equation}
\vec{\mathcal{T}} = e_{p} \vec{r}_p + e_{n} \vec{r}_n,
\label{eq:E1full}
\end{equation}
%%%%%%
where $e_{p}$ ($e_{n}$) is the proton (neutron) E1 effective charge, and $\vec{r}_p$ ($\vec{r}_n$) is the 
coordinate operator of protons (neutrons).
We employ the usual values $e_{p}=e N/A$ and $e_{n}=-eZ/A$, so as to exclude the center-of-mass 
motion.  We next introduce the operator with three components of $\vec{\mathcal{T}}$,
%%%%%  eq.(4)
\begin{equation}
\mathcal{X} \,=\, \exp(i \, \zeta \, (\mathcal{T}_x + \mathcal{T}_y + \mathcal{T}_z) ),
\label{eq:xop}
\end{equation}
%%%%%%%%
where $\zeta$ is a real-number parameter. 
An appropriate common value of $\zeta$ is used in this work.  Note that a different combination of $\mathcal{T}_x$, $\mathcal{T}_y$ and $\mathcal{T}_z$ is equally valid in principle.  
We act this operator on the MCSM basis vectors for the ground state as
%%%%%%%  eq.(5)
\begin{equation}
|\, \xi^{(n)}  \rangle \, = \, \mathcal{X} \,|\, \phi^{(n)}  \rangle, \,\,\,\,n = 1, 2, ..., N_B.  
\label{eq:E1MCSMbasis}
\end{equation}
At the infinitesimal limit of $\zeta \rightarrow 0$, this action produces the E1 transition from $|\, \phi^{(n)} \rangle$.  
Because $|\, \phi^{(n)} \rangle$ is a Slater determinant, the state $|\, \xi^{(n)}  \rangle$ is another Slater determinant
for any value of $\zeta$.  Namely, eq.~(\ref{eq:E1MCSMbasis}) produces another set of $N_B$ Slater determinants with E1 transition character.  We then diagonalize the Hamiltonian in the subspace spanned by the basis vectors $| \, \xi^{(n)}  \rangle$ (n = 1, ..., $N_B$) with the projection onto $J^{\pi}= 1^{-}$.  
%the basis vectors for the E1 excitation from the ground state.  
From the construction, it is expected that the E1 excitation may be described to a certain extent, and we shall discuss how good the approximation is.      
In the usual MCSM, basis vectors are optimized so as to lower the corresponding energy, but in the present scheme, the basis vectors are created by the E1 transitions and the resulting basis vectors make sense only for the specific initial state. 

%%%  Figure 2   
\begin{figure}[tb]
  \includegraphics[width=8.2cm]{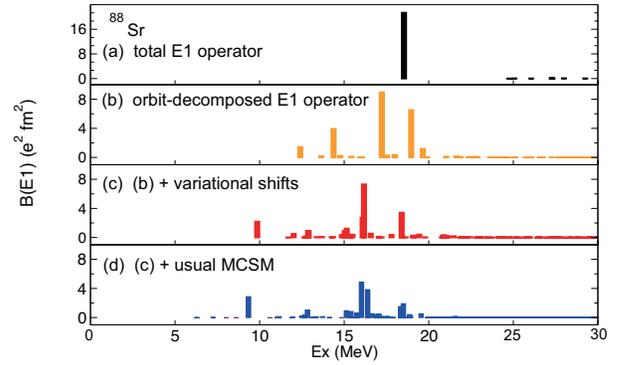}
  \caption{(Color online)  B(E1) values obtained in set (a), (b), (c), and (d) calculations.
  MCSM basis vectors are generated  by (a) E1 operator, (b) orbit-decomposed E1 operators, 
  (c) (b)+variational refinements, (d) (c)+usual MCSM basis vectors.  See the text.    }
  \label{fig:4step}
\end{figure}

Before looking into the actual calculations, we briefly describe the Hamiltonian used.  
The effective nucleon-nucleon interaction is expressed in terms of so-called two-body matrix elements (TBMEs) 
between two-nucleon states.
% where the nucleons are in assigned single-particle orbits and are coupled to given angular momentum $J$ and isospin $T$.  
%The actual numerical calculations are extremely heavy.  In order to save computer resources, 
We start with existing sets of TBMEs, and modify them in a simple way. 
The $pf$-shell part is taken from a standard SM interaction, GXPF1A\cite{honma2005}.  The interaction involving the 1$g_{9/2}$ and the $pf$-shell orbits except for 1$f_{7/2}$ is taken from another standard SM interaction, JUN45\cite{jun45}.  
The SNBG3 SM interaction\cite{snbg3} is used for $T$=1 TBMEs involving 1$g_{7/2}$, 2$d_{5/2,3/2}$, 3$s_{1/2}$ and 1$h_{11/2}$.    Note that these interactions were constructed by performing empirical fits to microscopically derived 
TBMEs  \cite{honma2005,jun45,snbg3,hjensen1995}.   
%Note that the JUN45 (SNBG3) set was obtained, in the corresponding model space, 
%by modifying TBMEs calculated from microscopic interactions called G-matrix \cite{hjensen1995} based on the CD-Bonn \cite{machleidt1989} (N3LO \cite{entem2003}) interaction, so as to better reproduce relevant experimental energies \cite{jun45,snbg3}.  
The $V_{{\rm MU}}$ interaction \cite{vmu} is taken for the rest of TBMEs. 
It consists of the central part given by a Gaussian 
function in addition to the $\pi$- and $\rho$-meson exchange 
tensor force \cite{vmu}.  The parameters of this Gaussian function were fixed from monopole 
components of known SM interactions \cite{vmu}.
The effective interaction thus constructed appears to be too strong, and we reduce central-force parts or some TBMEs of similar nature.   Such reduction factors are natural, because the original TBMEs were obtained in smaller
model spaces but the present space is much bigger.  We determined this reduction factor so as to reproduce the excitation energy of the 2$^+_1$ state of the $^{88}$Sr nucleus, and use this Hamiltonian also for other nuclei.  
The SNBG3 set was not reduced simply because the present calculation is insensitive to it.  
The single-particle energies are determined to be consistent with the TBMEs (or Woods-Saxon potential for the highest ones), except for the 1$f_{7/2}$ orbit.  Its energy cannot be fixed by existing sets of the interaction, and 
is determined so that its hole energy is about 4 MeV above the 2$p_{1/2}$ hole energy in $^{89}$Y, referring to its observed levels \cite{nudat2}.  These parameters can be fine tuned in future work.  The spurious center-of-mass motion is pushed up to higher energy in a standard way by giving extra positive energy to this motion\cite{gloeckner1974}, and is confirmed to be well separated.   

We first calculate the wave function of the ground state with the usual MCSM with this Hamiltonian, with $N_B =15$ (see eq.~(\ref{eq:eigenstate})).   We then generate the MCSM basis vectors for the $1^{-}$ states by eq.(\ref{eq:E1MCSMbasis}), and diagonalize the same Hamiltonian with those basis vectors projected onto $J^{\pi}= 1^{-}$.  The E1 transition strength is quantified in terms of the B(E1) value \cite{bohr_mottelson,blatt_weisskopf}, and Figure~\ref{fig:4step}(a) shows the B(E1) strength from the ground state to the eigenstates thus obtained.  One sees a high peak around $E_x$=18 MeV as well as small peaks at higher energies. This set of the calculation is referred to as set (a), hereafter.  

\begin{figure}[tb]
  \includegraphics[width=8cm]{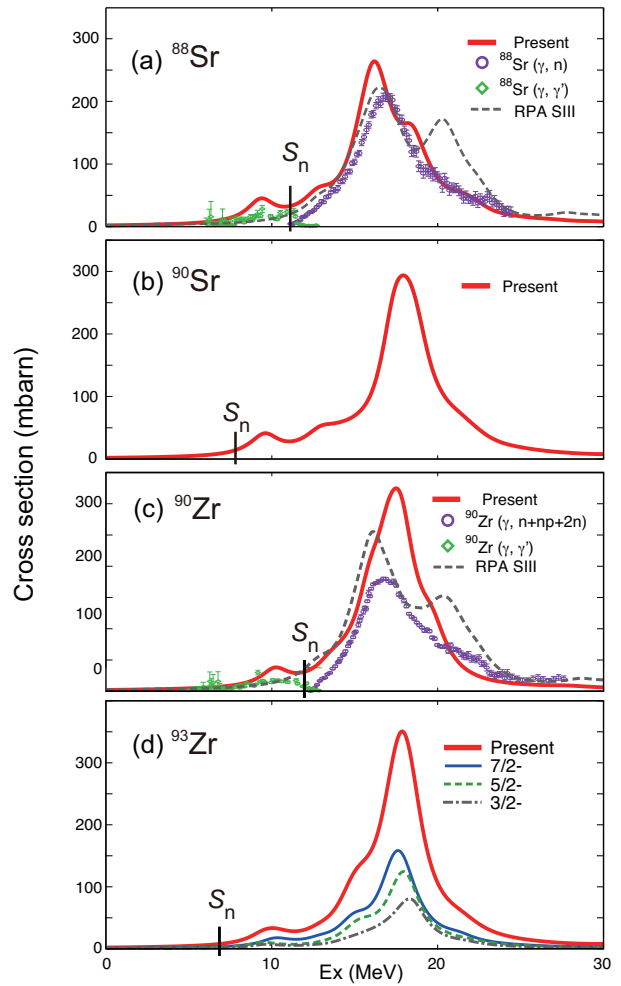}
  \caption{(Color online)  Cross section for the $\gamma$-induced E1 excitation for 
  (a) $^{88}$Sr, (b) $^{90}$Sr, (c) $^{90}$Zr, and (d) $^{93}$Zr. 
  Experimental data are taken from \cite{88Sr_gn,88Sr_gg,90Sr_gn,90Sr_gg}. 
  The RPA calculation with the SIII interaction \cite{beiner1975} is included in (a) and (c). 
  In (d), Present means the total value. }
  \label{fig:spec}
\end{figure}

Figure~\ref{fig:spec}(a) displays measured photoabsorption cross sections with a large peak around $E_x$=17 MeV.
Thus, this prescription is quite promising already in its lowest order.  In order to have various properties of the E1 excitations, we improve the method by incorporating actual shell structure.   The $\mathcal{X}$ operator is decomposed into individual combinations of the single-particle orbits $j$ and $j'$ as
%%%%%%  eq. (6)
\begin{equation}
\mathcal{X}_{j,\,j'} \,=\, \exp{ \Bigl( \, i \, \zeta \, (\mathcal{T}_{j,\,j'; \,x}+\mathcal{T}_{j,\,j'; \,y}+\mathcal{T}_{j,\,j'; \,z}) \,\Bigr) },
\label{eq:xop_orbit}
\end{equation}
where the operator $\mathcal{T}_{j,\,j'; \,x,y,z}$ carries the E1 process from the orbit $j'$ to $j$, with $\vec{\mathcal{T}}$ equal to the sum of all possible $\vec{\mathcal{T}}_{j,\,j'}$'s.  Likewise, $\mathcal{X}_{j,\,j'}$ is for this specific pair of the orbits $j$ and $j'$.
Note that $\mathcal{X}_{j,\,j'}$ transforms a Slater determinant to another single Slater determinant.
This decomposition is rather essential because different combinations of $j$ and $j'$ can connect to different Slater determinants from the common initial Slater determinant, and their proper superposition should give a more refined  description.   %Without this decomposition, an average effect may be seen.

%%%  Figure 3   

In the present case, there are 18 combinations: $j$ =1$g_{9/2}$ and $j$'=1$f_{7/2}$, {\it etc}.
We take all of them, and create a subspace spanned by 15 (=$N_B$) $\times$ 18 = 270 basis vectors.
The results are called set (b), with their $B(E1)$ values shown in Figure~\ref{fig:4step}(b).
We now see a more refined E1 spectrum.

The set (b) is improved further, by adding more basis vectors.  We choose $N_{X}$ most important 
$\mathcal{X}_{j,\,j'}$ operators based on their contributions to the B(E1) values.   Actually dropped ones contribute
much less than the selected ones.  We take $N_{X}$=10 here.  Suppose $\mathcal{X}_{\ell,\,\ell'}$ being one of the selected terms, we perform a variational shift 
%%%%%%%  eq.(7)
\begin{equation}
\mathcal{V} \left[ \mathcal{X}_{\ell,\,\ell'} \,\phi^{(n)} \right],
\label{eq:set3}
\end{equation}
where $\mathcal{V}  \left[ \, \right]$ stands for a variational functional for the Slater determinant in the brackets, 
in its vicinity and under the presence of the other basis vectors.  We carry out this for the $N_B \times N_X$ combinations, adding 15$\times$10=150 more basis vectors and yielding 420 in total.  
%This variational process is repeated once more, adding 150 basis vectors in total. Not done !
Figure~\ref{fig:4step}(c) displays such set (c) result, where major peaks are lowered from set (b) as 
expected from the variation.  We note also that a peak arises around Ex=10 MeV, as two peaks around 12 and 14 MeV in panel (b) disappear.    

For set (d), we add another 240 basis vectors generated with the usual MCSM procedure with the $J^{\pi}= 1^{-}$ projection, as shown in Figure~\ref{fig:4step}(d).  The major peaks remain almost unchanged from set (c), while some small peaks appear in the low energy region, also certain rearrangements among nearby peaks occur.
The minor change between the sets (c) and (d) suggests the validity of the present scheme.

The validity of the calculation can be examined in an independent manner.   The E1 sum from the ground state can be
calculated by $\langle \Psi \,({\bf D})\, |( \vec{\mathcal{T}} \cdot \vec{\mathcal{T}} ) | \, \Psi ({\bf D}) \rangle$ with $( \cdot )$ being a scalar product.  The same quantity can be evaluated by summing the B(E1) values obtained in the sets (b) or (c) with the value reaching 90\% of the sum, which implies that most of the basis vectors needed to account for the E1 excitation strength are generated.  If one needs to improve the description, we can have two values of $\zeta$, which makes the size of the calculation twice.  This would not make much sense at moment.

The cross section for the $\gamma$-induced E1 excitations is shown in Fig.~\ref{fig:spec}, where the B(E1) values are transformed into the cross section by
%%% eq(8)
\begin{eqnarray}
\sigma(E)
=
   \frac{16\pi^{3}}{9} \frac{e^{2}}{\hbar c} \sum_{J^{f}_{n}} \frac{1}{\pi}  
   \frac{\Gamma/2}{(E - E_{x}(J^{f}_{n}))^2+(\Gamma/2)^2} \nonumber\\ \times \, E_{x}(J^{f}_{n}) \times 
   B(E1; J^{i} \rightarrow J^{f}_{n} ),
\label{eq:xsection}
\end{eqnarray}
where $J^{i}$ ($J^{f}_{n}$) stands for the spin/parity of the initial ($n$-th final) state,  $E_{x}(J^{f}_{n})$ denotes the excitation energy of the $n$-th final state, $\Gamma$ is the width used in the Lorentzian smearing.
Figure~\ref{fig:spec} depicts the photoabsorption cross section of $^{88,90}$Sr and $^{90,93}$Zr, where 
the present result is obtained from the set (d).  Experimental data are available for $^{88}$Sr and $^{90}$Zr \cite{88Sr_gn,88Sr_gg,90Sr_gn,90Sr_gg}, compared to which the present result shows rather good agreement.  
%We note that the parameters are fitted for $^{88}$Sr, and the other three cases are predictions.     
Large peaks at $E_x$=16-18 MeV correspond to the GDR.  The present results resemble the spectra obtained by the RPA calculation with the Skyrme SIII model \cite{beiner1975}, apart from an additional peak at higher energy with SIII.   The present calculation shows a small peak around 9 MeV in each case, called the PDR for brevity.  The experimental data are included in the figure, which show similar enlargements of the cross section, or PDR.  These experimental data are discussed in \cite{beard2012} in relation to the astrophysical implication and to the transmutation of such  long-lived fission products.  Those PDRs are seen by the $(\gamma, \gamma')$ reaction, where  excited states populated by the reaction decay into many states.  Some of the decays are undetected usually, resulting in the underestimate of the cross section, while some corrections were made in \cite{88Sr_gg}.   The present calculation predicts the energy of those peaks rather well.   The experimental value of the cross section is below the present calculation and above the RPA result, urging further investigation theoretically and experimentally.  
We point out that the PDR appears in all of the four nuclei, while the neutron threshold moves downwards as $N$ increases.  In this case, once the PDR is excited, the neutron emission occurs, leading to another nucleus.  Considering that 
$^{90}$Sr and $^{93}$Zr are long-lived fission products in nuclear waste, it should be very helpful to society  if they could be eliminated somehow by utilizing these excitations.  The relevance to the 
astrophysics is also important  \cite{beard2012}.   We stress that the odd-mass nucleus $^{93}$Zr can be described as well as other even-mass nuclei.  

% Figure 4
\begin{figure}[tb]
\includegraphics[width=8.6cm]{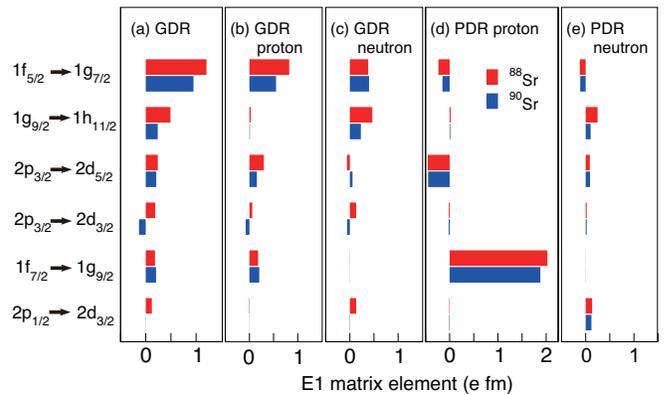}
\caption{ (Color online)
E1 matrix elements, including proton and neutron decompositions, for $^{88,90}$Sr calculated for discrete states before the Lorentzian smearing.
Those states are taken from two major peaks of the GDR and PDR.    
}
\label{fig:E1mat}
\end{figure}

%It is of interest to see the contributions of individual transitions.   
Figure~\ref{fig:E1mat} displays E1 matrix elements (including effective charges) for six pairs of the initial and final orbits for $^{88,90}$Sr.  The panel (a) shows the total matrix elements for the highest discrete peak of the GDR before the Lorentzian smearing, while panels (b) and (c) display its proton and neutron contents.  Figure~\ref{fig:E1mat}(d,e) indicates the same quantities for the PDR peak around $E_x$=9 MeV.   
The GDR exhibits coherent contributions between proton and neutron, and also among many orbital pairs.  
On the other side, the PDR has a single major contribution from the proton transition 1$f_{7/2} \rightarrow 1g_{9/2}$, 
suggesting its single-particle character.    It is quite intriguing that the PDR (or something similar) appears in nuclei 
without much neutron excess due to proton excitations, 
in contrast to the frequently conceived picture involving neutron excitations, {\it e.g.}, \cite{litvinova2008,inakura2011,ebata2014}.  This feature should be studied more.

  %%%%%%%%%%%%%%%%%%%%%%%%%%%%%%%%%%%%%%%%%%%%%%%%%%%% 
 
In summary, we presented the novel method to calculate the excitation spectrum on top of a given MCSM eigenstate with full of correlations. This method is innovative in the sense that basis vectors important for a given excitation mode are generated, and the spectrum is obtained by diagonalizing the Hamiltonian with them.  The method works with general SM interactions and is applicable to various modes such as E1, M1, isoscalar E1, E2, as well as Gamow-Teller.  The applications to the E1 excitation from the ground states of $^{88,90}$Sr and $^{90,93}$Zr are presented as proof-of-principle, accounting for 90\% of the total B(E1) value and for the GDR and PDR peaks.  This method can be applied to even-even, odd-A and odd-odd nuclei, with spherical, deformed and transitional shapes all inclusive, and also to the {\it ab initio} MCSM calculation\cite{tabe2012}.
It is thus complementary to other methods such as various kinds of RPA or the Coupled-Cluster approach \cite{bacca2013,bacca2014}.
The position of the PDR can be very important for the transmutation of the long-lived fission products in  nuclear waste, and it would be nice if the present method can be of some use.

%\end{acknowledgements}

\end{document}